\documentclass{ws-ijmpa}
\usepackage[super,compress]{cite}
\usepackage{graphics}
\usepackage{dcolumn}
\usepackage{bm}
\usepackage{mathrsfs}
\usepackage{slashed}
\usepackage{epsfig}
\usepackage{amsfonts}
\usepackage{amsmath}
\usepackage{amssymb}
\def\beq{\begin{equation}}
\def\eeq{\end{equation}}
\def\bea{\begin{eqnarray}}
\def\eea{\end{eqnarray}}
\def\nn{\nonumber}
\def\q{{\mathbf q}}

\def\A{{\mathbf A }}
\def\B{{\mathbf B }}
\def\E{{\mathbf E }}
\begin{document}

\markboth{G. Krein, G. Menezes and N. F. Svaiter}
{Markovian versus non-Markovian stochastic quantization
of a complex-action model}

\catchline{}{}{}{}{}

\title{MARKOVIAN VERSUS NON-MARKOVIAN STOCHASTIC QUANTIZATION OF A COMPLEX-ACTION MODEL}
\author{G. KREIN}
\address{Instituto de F\'\i sica Te\'orica, Universidade Estadual Paulista\\
Rua Dr. Bento Teobaldo Ferraz 271 - Bloco II \\ S\~ao Paulo, SP 01140-070, Brazil
\\ gkrein@ift.unesp.br}
\author{G. MENEZES}
\address{Instituto de F\'\i sica Te\'orica, Universidade Estadual Paulista\\
Rua Dr. Bento Teobaldo Ferraz 271 - Bloco II \\ S\~ao Paulo, SP 01140-070, Brazil
\\ gsm@ift.unesp.br}
\author{N. F. SVAITER}
\address{Centro Brasileiro de Pesquisas F\'{\i}sicas, Rua Dr. Xavier
Sigaud 150\\ Rio de Janeiro, RJ 22290-180, Brazil
\\ nfuxsvai@cbpf.br}

\maketitle

\begin{abstract}
We analyze the Markovian and non-Markovian stochastic quantization methods
for a complex action quantum mechanical model analog to a Maxwell-Chern-Simons
eletrodynamics in Weyl gauge. We show through analytical methods convergence to
the correct equilibrium state for both methods. Introduction of a memory kernel
generates a non-Markovian process which has the effect of slowing down oscillations
that arise in the Langevin-time evolution toward equilibrium of complex action
problems. This feature of non-Markovian stochastic quantization might be beneficial
in large scale numerical simulations of complex action field theories on a lattice.
\keywords{Stochastic Quantization; Complex Actions; Topological Quantum Mechanics}
\end{abstract}

\ccode{PACS Nos.: 03.70+k, 05.10.Gg, 11.10.-z}

\section{Introduction}
\label{sec:intro}

Recent years have witnessed a vigorous revival~\cite{Berges:2005yt,Berges:2006xc,Berges:2007nr,Aarts:2008rr,Aarts:2008wh,Aarts:2009dg,Aarts:2009hn,Aarts:2010gr,Aarts:2010aq,Aarts:2011zn} of the method of stochastic quantization~\cite{Parisi:1980ys}
of systems with a complex action~\cite{Klauder1,Klauder2,Klauder3,Parisi:1984cs}. The revival comes after
drawbacks of the method, pointed out long ago~\cite{KlauPeter,AmbFlensPeter,Ambjorn:1985iw}
related to lack of convergence or convergence to a wrong limit of solutions of the associated
Langevin equations. The interest in stochastic quantization of complex actions is driven
mostly by the pressing need of simulation techniques to study the phase diagram of
quantum chromodynamics (QCD) at finite temperature and baryon chemical potential. The
traditional Monte Carlo methods widely used in studies of the hadron spectrum, based on
importance sampling with a Boltzmann weight given in terms of the real, positive Euclidean
action of QCD~\cite{Gattringer:2010zz}, is inapplicable to problems with a baryon chemical
potential because the action is complex. This difficulty is not exclusive to QCD, it
also occurs in several other physics problems; notorious examples are problems of cold atoms
and strongly-correlated electrons in condensed matter physics. The main difficulty with a
complex-action Langevin simulation can be better posed in terms of the stationary solutions
of the associated Fokker-Planck equation~\cite{Damgaard:1987rr}: whereas in the case of a real
action the Boltzmann weight of the Euclidean path integral can be shown to be given by the
stationary solution of the Fokker-Planck equation, such a proof is still lacking for a complex
action. Pragmatically, however, the correctness of a given complex-action Langevin simulation
can be assessed to some extent with the use of a set of rather simple and general
criteria~\cite{Aarts:2011ax} that calculated observables must satisfy.

The recent renewed optimism with stochastic quantization of complex actions has grown from
robust evidence that problems with instabilities and incorrect convergence of solutions of the
Langevin field equations can be controlled by choosing a small enough Langevin
step-size~\cite{Berges:2005yt} and also with the use of more elaborate algorithms, like
of adaptive step-size and of higher order~\cite{Aarts:2009dg}. A complex action in general
introduces oscillatory behavior in the time evolution toward equilibrium of the solutions of
the Langevin equation; depending on the problem, the oscillations become irregular and of high
frequency. High-frequency oscillations are the main reason for the need of smaller step-sizes.
With this in mind, in the present paper we advocate the use of non-Markovian stochastic
quantization for complex actions. Non-Markovian stochastic quantization amounts to a
modification of the Langevin equation by introducing a memory kernel and use of colored noise
according to the fluctuation and dissipation theorem~\cite{menezes2,menezes3,menezes4}.
A judiciously chosen memory kernel can soften considerably the oscillatory behavior induced by
a complex action and hence larger step-sizes can be used to sample the time evolution. We illustrate
this by making use of a simple quantum mechanical model with a complex action that is soluble
analytically. Specifically, we employ a topological quantum mechanical model~\cite{Dunne:1989hv}
which is analogous to the three-dimensional topologically massive, Chern-Simons electrodynamics
in the Weyl gauge~\cite{Deser:1982vy}.

Although our primary interest in the quantum mechanical model is non-Markovian
stochastic quantization, we emphasize that topological actions find interesting applications
in several situations of physical interest. For instance, coupling of Maxwell and Chern-Simons
Lagrangians in $2+1$ dimensions yields a different form of gauge field mass generation, known
in the literature as a topologically massive gauge theory~\cite{Deser:1981wh}. In addition,
it has been argued that topological Chern-Simons fields may play an important role in the
three-dimensional dynamics in planar condensed-matter settings. As discussed in Refs.~\refcite{q1,q2,q3},
such topological field configurations are present in models for the fractional quantum Hall effect
which encompasses quasiparticles with magnetic fluxes attached to charged particles.
More recently, models with Chern-Simons terms were also employed in the study of the pseudogap
phase in High-$T_C$ superconductors~\cite{marche}. Further discussions on peculiar features on
the quantization of topological field theories can be found in Ref.~\refcite{Menezes:2006pe}.

The organization of the paper is as follows. In section~\ref{sec:model}, we discuss
the topological model we study in the present paper. In section~\ref{sec:mark} we present the
Markovian stochastic quantization of the model and show that the associated Langevin equation
converges to the correct limit. The non-Markovian stochastic quantization of the same model
is discussed in section~\ref{sec:nmark}. Explicit numerical solutions are presented in
section~\ref{sec:equil} for both Markovian and non-Markovian processes. Conclusions and
perspectives are presented section~\ref{sec:concl}. The paper also includes an
Appendix, where we present details of some lengthy derivations.

\section{The model}
\label{sec:model}

The topological model we consider describes the motion of a particle of mass $m$ and
electrical charge $e$ in external electromagnetic fields. The external fields give rise
to Lorentz forces and the motion of the particle is governed by the standard
Lagrangian~\cite{Dunne:1989hv}:
\beq L =  \frac{m}{2} {\dot \q}^2 + \frac{e}{c} \dot\q \cdot \A(\q)
- eV(\q) , \label{lag-gen} \eeq
where $\q = (q^1,q^2)$ is the only dynamical variable of the system,
and $\A(\q) =(A^1(\q),A^2(\q))$ and $V(\q)$ are the external vector
and scalar electromagnetic potentials; the magnetic and electric
fields are given by $\B = \nabla\times\A$ and $\E = - \nabla V$. The
model is exactly solvable, classically and quantum-mechanically, for
a constant magnetic field~$B$, $A^i = - \epsilon^{ij} q^j B/2$, and
a quadratic scalar potential $V(\q)~=~k \q^2/2 $ -- summation over
repeated indices is implied. In this case, Eq.~(\ref{lag-gen})
becomes
\beq L =  \frac{m}{2} {\dot \q}^2 + \frac{eB}{2c} \q \times \dot\q -
\frac{e k}{2}\q^2 .
\label{lag}
\eeq
As mentioned in the Introduction, this Lagrangian is analogous to
the Lagrangian density of three-dimensional,
topologically massive electrodynamics in the $A^0 = 0$ gauge:
\beq {\cal L} = \frac{1}{2}\dot\A^2 + \frac{\mu}{2} \dot\A\times\A -
\frac{1}{2}(\nabla\times\A)^2 .
\label{lag-CS}
\eeq
Here, $\A$ is the only dynamical variable of the problem. The formal
correspondence between $L$ and ${\cal L}$ is such that the kinetic
and potential terms $m {\dot \q}^2/2$ and $-e k \q^2/2$ are analogous
to $\dot\A^2/2$ and $-(\nabla\times\A)^2/2$ respectively, and the term
corresponding to the Lorentz force $e B \q \times \dot\q /2 c$ is
analogous to the Chern-Simmons term $\mu \dot\A\times\A /2$. Henceforth,
we set $c$ and $e$ to unity.

The Hamiltonian corresponding to the Lagrangian of Eq.~(\ref{lag})
is given by
\beq
H = \frac{1}{2m} \biggl(p^{i} + \frac{B}{2}\epsilon^{ij} q^j\biggr)
\biggl(p^{i} + \frac{B}{2}\epsilon^{ik} q^k\biggr) + \frac{k}{2}q^{i}q^{i},
\eeq
with
\beq
p^{i}(t) = \frac{\partial L}{\partial \dot{q}^i(t)} = m \dot{q}^i(t)
- \frac{B}{2}\epsilon^{ij} q^j(t).
\label{pq}
\eeq
The Hamiltonian can be brought to diagonal form~\cite{Dunne:1989hv}:
\beq
H = \frac{1}{2}\left(p^2_+ + \omega^2_+ q^2_+ \right) + \frac{1}{2} \left( p^2_- +
\omega^2_- q^2_-\right),
\label{H_diag}
\eeq
where $(p_{\pm},q_{\pm})$ are canonical variables given in terms of the
original $(p^i, q^i)$ as
\bea
&&p_{\pm} = \biggl(\frac{\omega_{\pm}}{2 m\,\Omega}\biggr)^{1/2}p^1
\pm \biggl(\frac{m\,\Omega\omega_{\pm}}{2}\biggr)^{1/2}q^2 \nonumber\\
&&q_{\pm} = \biggl(\frac{m\, \Omega}{2\omega_{\pm}}\biggr)^{1/2}q^1
\mp \biggl(\frac{1}{2m\,\Omega\omega_{\pm}}\biggr)^{1/2}p^2,
\label{can-var}
\eea
with the frequencies $\omega_\pm$ given by
\beq
\omega_{\pm} = \Omega \pm \frac{B}{2m},
\label{omega-pm}
\eeq
where
\beq
\Omega = \biggl(\frac{B^2}{4m^2} + \frac{k}{m}\biggr)^{1/2}.
\label{Omega}
\eeq

Path integral quantization of the model proceeds via the probability
amplitude $Z$ of finding the particle at position $\q$ at time
$t$, when one knows that it was located at point $\q_0$ at time $t_0$
-- for simplicity of presentation, we set $\q_0 = 0$ and $t_0 = 0$.
An analytical continuation of $Z$ to imaginary time $t
\rightarrow -\imath \,t$ leads to the following path integral representation
of~$Z$:
\beq
Z = \int {\cal D}q(t) \, e^{ -  S[\q] } ,
\label{Z}
\eeq
where the action $S[\q]$ corresponding to the Lagrangian in
Eq.~(\ref{lag}) is complex and given by
\beq
S[\q] = \int dt \Bigl[ \frac{m}{2} \, \dot{q}^i(t) \dot{q}^i(t)
- \imath \frac{B}{2}\epsilon^{ij} q^i(t) \dot{q}^j(t)  +
\frac{k}{2} q^i(t) q^i(t) \Bigr]. \label{S-eucl}
\eeq
Equivalently, using the the coordinates and velocities $(q_\pm,\dot q_\pm)$
corresponding to the canonical variables $(q_\pm,p_\pm)$ given in Eq.~(\ref{can-var}),
the action is given by
\beq
S[q_{\pm}] = \int dt \Bigl[ \frac{1}{2} \, \bigl(\dot{q}^2_{+}(t) + \dot{q}^2_{-}(t)\bigr)
+ \frac{1}{2}\bigl( \omega^2_{+}q^2_{+}(t) + \omega^2_{-}q^2_{-}(t) \bigr)\Bigr].
\label{S2-eucl}
\eeq
From the correlation function
\beq \Delta^{ij}(t,t') = \frac{1}{Z[q]} \int {\cal D}q(t) \, q^i(t)
q^j(t') \, e^{ - S[q] },
\label{corr}
\eeq
one can obtain the energy gap between the ground-
and the first excited-state from the large-time $|t-t'| \rightarrow \infty$
falloff of $\Delta^{ij}(t,t')$. Specifically, for the present model:
\beq
\lim_{|t-t'|\rightarrow \infty} \Delta^{\pm \, \pm}(t,t') \sim
 e^{- \Delta E_{\pm} |t-t'|},
\label{asympt}
\eeq
where $\Delta E_{\pm}$ is given by
\beq \Delta E_{\pm} = \omega_\pm = \sqrt{\frac{B^2}{4m^2} + \frac{k}{m}} \pm
\frac{B}{2m}.\label{dE1} \eeq
For the so called reduced theory considered in Ref.~\cite{Dunne:1989hv}, for which
$m = 0$, one can show that
\beq
\Delta E_{\pm} = \pm\frac{k}{B}.
\label{dE2}
\eeq
%

%
\section{Markovian stochastic quantization}
\label{sec:mark}

Markovian stochastic quantization (MSQ) is based on a Langevin equation of
the form
\begin{equation}
\frac{\partial}{\partial\tau}q^{i}(\tau,t) = -
\frac{\delta\,S}{\delta\,q^{i}(\tau,t)} + \eta^{i}(\tau,t),
\label{lang-mark}
\end{equation}
where time $\tau$ is a fictitious time variable, $S$ is the action given by
Eq. (\ref{S-eucl}), and $\eta^i(\tau,t)$ is postulated to satisfy
\bea
&& \langle\,\eta^{i}(\tau,t)\,\rangle _{\eta} = 0,
\label{noise-zero}\\
&& \langle\,\eta^{i}(\tau,t)\,\eta^{j}(\tau',t')\,\rangle _{\eta}
= 2 \delta^{ij} \delta(\tau-\tau') \delta(t-t').
\label{noise-white}
\eea
Here, $\langle \cdots \rangle_\eta$ means ensemble average over
noise realizations. Expectation values $\langle O[q(t)] \rangle$
of quantum mechanical operators $O[q(t)]$ are obtained as ensemble
averages of the functions $O[q(\tau,t)]$ in the $\tau \rightarrow \infty$
limit. In particular, the correlation function $\Delta^{ij}(t,t')$ defined
in Eq.~(\ref{corr}) is obtained as
\beq
\Delta^{ij}(t,t') = \lim_{\tau'=\tau \rightarrow \infty} \langle q^i(\tau,t)
q^j(\tau',t') \rangle_\eta ,
\label{corr-sq}
\eeq
where the $q(\tau,t)$ are solutions of the Langevin equation in
Eq.~(\ref{lang-mark}).

Solutions of the Langevin equation in Eq.~(\ref{lang-mark}) can be obtained
as follows. Since the equation is linear, it can be solved using Fourier
transforms in $t$ for $q^i(\tau,t)$ and $\eta^i(\tau,t)$:
\beq
\left(\begin{array}{c}
  q^i(\tau,t)  \\
  \eta^i(\tau,t)\\
\end{array}\right) = \int^{+\infty}_{-\infty} d\omega  \; e^{\imath\omega t}
\left(\begin{array}{c}
  q^i(\tau,\omega)  \\
  \eta^i(\tau,\omega)\\
\end{array}\right) .
\eeq
Using these in Eqs.~(\ref{lang-mark}) and (\ref{noise-white}), one
obtains
\beq
\frac{\partial}{\partial\tau}q^{i}(\tau,\omega) = -
\left[(m\omega^2 + k) \delta^{ij} +
B\varepsilon^{ij}\omega \right]\,q^j(\tau,\omega)
+ \eta^{i}(\tau,\omega) ,
\label{lang-w}
\eeq
and
\begin{equation}
\langle\,\eta^{i}(\tau,\omega)\,\eta^{j}(\tau',\omega')\,\rangle
_{\eta} = 4\pi \delta^{ij} \delta(\tau-\tau')
\delta(\omega + \omega').
\label{eta-eta_w}
\end{equation}
It is instructive to represent the solution of Eq.~(\ref{lang-w})
in terms of a retarded matrix-valued Green's function with elements
$g^{ij}(\tau,\omega)$:
\begin{equation}
q^i(\tau,\omega) =
\int^{\tau}_0 d\tau' \, g^{ij}(\tau-\tau',\omega) \,
\eta^{j}(\tau',\omega),
\label{sol-w}
\end{equation}
where we assumed $q^i(0,\omega) = 0$, with $g^{ij}(\tau,\omega)$
obeying the differential equation:
\beq
\frac{\partial}{\partial\tau} g^{ij}(\tau,\omega) =
- \left[(m\omega^2 + k) \delta^{ik} +
B\varepsilon^{ik}\omega \right] \,g^{kj}(\tau,\omega)  \nonumber\\
+ \delta^{ij} \, \delta(\tau) .
\eeq
The solution of this equation is:
\beq
g^{ij}(\tau,\omega) = \theta(\tau) \, \left[\delta^{ij} \cos(B\omega\tau) -
\varepsilon^{ij} \sin(B\omega\tau) \right] \, e^{-(m\omega^2 + k)\tau},
\label{mg}
\eeq
where $\theta(\tau)$ is the usual step function. Plainly, $g^{ij}(\tau,\omega)$
is oscillatory in $\tau$ because $B \neq 0$ - recall that $B \neq 0$ implies a
complex Euclidean action, Eq.~(\ref{S-eucl}). The oscillatory behavior of
the Langevin evolution is a generic feature of a complex action problem, and is
the main cause of instabilities or of convergence to wrong limits in numerical
integration procedures.

For $B$ purely imaginary, the trigonometric functions in Eq.~(\ref{mg})
lead to an exponentially growing factor $e^{|B|\omega\tau}$. Also, for the
purely topological theory, i.e. $m = k = 0$, one immediately sees that there will
be no large-$\tau$ limit for the solution $q_i(\tau,\omega)$ and their correlation
functions. In other words, the stochastic process described by the Langevin equation
in Eq.~(\ref{lang-mark}) never approaches an equilibrium solution for
the purely topological theory. These quantities can only be set to zero in the
equilibrium results. Still, we are free to take one of the parameters $m,k$ as zero
while maintaining the other finite.

Using standard methods, one can calculate easily the two-point correlation
function $\langle q^i(\tau,\omega)q^j(\tau,\omega')\rangle_{\eta}$ in the
large-$\tau$ limit. Using Eqs.~(\ref{eta-eta_w}), (\ref{sol-w}) and (\ref{mg}),
we obtain for the two-point correlation function
\beq
\Delta^{ij}(\omega,\omega') = \lim_{\tau \rightarrow \infty} \, \langle q^i(\tau,\omega)
q^j(\tau,\omega')\rangle_{\eta}
= \frac{2\pi\delta(\omega + \omega')}{p^4 + \omega^2B^2} \,
(\delta^{ij}p^2- \varepsilon^{ij} \omega B ),
\label{mcorr}
\eeq
where $p^2 = m\omega^2 + k$. From this, for the purely
topological theory one obtains
\begin{equation}
\Delta^{ij}(\omega,\omega') = -\frac{2\pi\delta(\omega + \omega')}{\omega B} \, \varepsilon^{ij}.
\end{equation}

As it stands, the result in Eq.~(\ref{mcorr}) indicates that the considered
stochastic process converges and the natural question is that if the converged
result is the correct one. The question can be answered by checking the asymptotic
behavior for the inverse Fourier transform of the two-point correlation function.
As remarked at the end of the previous section, the first energy gap can be
extracted from the large relative time behavior of the two-point correlation
function. It is easy to prove that Markovian stochastic quantization leads
to the correct limit given in Eq.~(\ref{asympt}). Since we are working in Euclidean
space, Eq.~(\ref{pq}) must be analytically continued to imaginary time. In Fourier
space:
\beq
p^{i}(\omega) = - m\omega q^{i}(\omega) - \frac{B}{2}\varepsilon^{ij}q^{j}(\omega),
\eeq
and hence
\beq
q_{\pm}(\omega) = m \left(\frac{1}{{2m\Omega\omega_{\pm}}}\right)^{1/2}
\left[\omega_{\mp}q^1(\omega) \pm \omega q^2(\omega) \right].
\eeq
With the help of Eq.~(\ref{mcorr}), one obtains
\bea
\Delta^{\pm\,\pm}(\omega,\omega') &=& \lim_{\tau \rightarrow \infty} \,
\langle q_{\pm}(\tau,\omega)q_{\pm}(\tau,\omega')\rangle_{\eta} \nn \\[0.3true cm]
&=& \frac{2\pi\delta(\omega + \omega')}{2m\Omega\omega_\pm(\omega^2 + \omega^2_{+})
(\omega^2 + \omega^2_{-})}\bigr[(\omega^2_{\mp} - \omega^2)p^2
+ 2 B \omega^2 \omega_{\mp}\bigl],
\label{corr-mark1}
\eea
where we used
\begin{equation}
p^4 + \omega^2B^2 = m^2\Biggl[\omega^4 + \omega^2\biggl(\frac{B^2}{m^2} + \frac{2 k}{m}\biggr)
+ \frac{k^2}{m^2}\Biggr]
= m^2 \left(\omega^2 + \omega^2_+\right)
\left(\omega^2 + \omega^2_-\right).
\eeq
Performing the inverse Fourier transforms of $q_{\pm}(\tau,\omega)$, one obtains
\begin{eqnarray}
\Delta^{\pm\,\pm}(t,t') &=& \frac{1}{2m\Omega\omega_+(\omega^2_{-} - \omega^2_{+})}
\biggl\{\pm \frac{e^{- \omega_{\pm} |t-t'|}}{2\omega_\pm}\Bigl[(\omega^2_{-}
+ \omega^2_{+})(-m\omega^2_{\pm} + k) \nonumber\\
&& \,\mp 2 B \omega_{\mp} \omega^2_{\pm}\Bigr]
 \mp \frac{e^{- \omega_{\mp} |t-t'|}}{2\omega_\mp}
\Bigl[2\omega^2_{\mp} (-m\omega^2_{\mp} + k) \mp 2 B \omega_{\mp}^3\Bigr] \biggr\}.
\label{fev1}
\end{eqnarray}
Using now the results:
\bea
&& 2\omega^2_{\pm} (-m\omega^2_{\pm} + k) \pm 2 B \omega_{\pm}^3
= \mp 2 B\omega^2_{\pm}(\pm B/2m + \Omega) \pm 2 B \omega_{\pm}^3 = 0,
\\
&& \frac{1}{\omega_\pm} \, \left[(\omega^2_{-} + \omega^2_{+})(-m\omega^2_{\pm} + k)
\mp 2 B \omega_{\mp} \omega^2_{\pm} \mp 2 B  \omega_{-} \omega_{+}\right]
= \mp 4 B \Omega^2,
\\
&& \omega^2_{-} - \omega^2_{+} = -\frac{2\Omega B}{m},
\eea
Eq.~(\ref{fev1}) can be cast in a simpler form as
\begin{eqnarray}
\Delta^{\pm\,\pm}(t,t') = \frac{e^{- \omega_{\pm} |t-t'|}}{2\omega_\pm} ,
\end{eqnarray}
which agrees with the result from the path integral calculation, Eq.~(\ref{asympt}).
We also note that for $m=0$, one obtains $\omega_{\pm}\equiv \pm \, {k}/{B}$, which yields
the first energy gaps in the reduced theory~\cite{Dunne:1989hv}.

To conclude this Section, we mention that Eq.~(\ref{lang-mark}) can be considered as the high-friction (or overdamped) limit of appropriate phase-space equations~\cite{horowitz1,horowitz2}. Ref.~\refcite{Luscher:2011qa} presents a recent discussion of problems with the continuum limit of second-order Langevin equations in numerical simulations of lattice field theories.

%
\section{Non-Markovian stochastic quantization}
\label{sec:nmark}

In the present section we consider the non-Markovian stochastic quantization (NMSQ)
of the model. NMSQ amounts to introduce a memory kernel $M_\Lambda(\tau-\tau')$ in
the Langevin equation as
\beq
\frac{\partial}{\partial\tau}q^{i}(\tau,t)
= -\int_{0}^{\tau}d\tau' \, M_\Lambda(\tau-\tau')
\frac{\delta S}{\delta q^{i}(\tau',t)} +  \eta^{i}(\tau,t),
\label{lang-mem}
\eeq
where $\Lambda$ is a parameter that controls the memory decay, such that
$M_\Lambda(\tau-\tau')\rightarrow\delta(\tau-\tau')$ as
$\Lambda\rightarrow\infty$, recovering the Markovian Langevin
equation of Eq.~(\ref{lang-mark}) in this limit. In order
to obtain the correct equilibrium distribution $\exp (-S[q])$, with the quadratic action $S[q]$
given by Eq.~(\ref{S-eucl}), one must impose the colored-noise
correlation:
\beq
\langle\,\eta^{i}(\tau,t)\,\eta^{j}(\tau',t')\,\rangle _{\eta}
= 2 \delta^{ij} M_\Lambda(\tau-\tau') \delta(t-t'),
\label{noise-color}
\eeq
instead of the white-noise form of Eq.~(\ref{noise-white}). An interesting situation with a non-Markovian approach was considered in Ref.~\refcite{craig:2002}. However in such a reference the author is concerned with the physical consequences associated with a non-Markovian friction, whereas here the generalized stochastic process is described by regarding a non-Markovian driving term in the Langevin equation.

As previously, performing a Fourier transform in $t$ for $q_i$ and
$\eta_i$ one obtains:
\begin{equation}
\frac{\partial}{\partial\tau}q_{i}(\tau,\omega) =-(p^2\delta_{ij} +
B\varepsilon_{ij}\omega)\,\int_{0}^{\tau}d\tau' \, M_\Lambda(\tau-\tau')
\,q_j(\tau',\,\omega)+\eta_{i}(\tau,\omega),
\label{nml}
\end{equation}
with
\begin{equation}
\langle\,\eta_{i}(\tau,\omega)\,\eta_{j}(\tau',\omega')\,\rangle
_{\eta}=4\pi\,\delta_{ij}M_\Lambda(|\tau-\tau'|)\,\delta(\omega +
\omega'), \label{nmeta}
\end{equation}
and, of course, $\langle\,\eta_{i}(\tau,\omega)\,\rangle=0$. The solution of
this Langevin can be written as
\begin{equation}
q_i(\tau,\omega) =
\int d\tau' \, G_{ij}(\tau-\tau',\omega) \, \eta_{j}(\tau',\omega),
\label{nmq}
\end{equation}
where we assumed $q_i(0,\omega) = 0$ and $G_{ij}$ is the retarded Green's
function obeying the differential equation
\begin{eqnarray}
\frac{\partial}{\partial\tau}G_{ij}(\tau,\omega) = - (p^2\delta_{ik}
+ B\varepsilon_{ik}\omega)
\,\int_{0}^{\tau} d\tau' \, M_\Lambda(\tau-\tau') G_{kj}(\tau',\,\omega) +
\delta_{ij}\,\delta(\tau). \label{nmg1}
\end{eqnarray}
This equation can be solved via the use of Laplace transformation. Formally,
one can write the solution as $G(\tau, \omega) = \Gamma(\tau,\omega)\theta(\tau)$,
with the $\Gamma$ matrix defined through the Laplace transform of its inverse:
\begin{equation}
\Gamma^{-1}_{ij}(z,\omega) = z \, \delta_{ij} +
D^{-1}_{ij}(\omega) \, M_\Lambda(z),
\label{gamma}
\end{equation}
with
\beq
D^{-1}_{ij}(\omega) = \delta_{ij} p^2 + \varepsilon_{ij}\, \omega  B ,
\label{D-1}
\eeq
and $M_\Lambda(z)$ is the Laplace transform of $M_\Lambda(\tau)$. For an exponential
kernel, $M_\Lambda(z)$ is given explicitly in Eq.~(\ref{Mz}); for such a kernel, the
$\Gamma$ matrix can be inverted analytically. As~outlined in the Appendix, after a
rather lengthy calculation one can write $\Gamma_{ij}(\tau,\omega)$ as
\begin{equation}
\Gamma_{ij}(\tau,\omega) = \delta_{ij}  I_{+}(\tau, \omega)
- \imath \, \varepsilon_{ij} I_{-}(\tau,\omega),
\label{nmg2p}
\end{equation}
where
\beq
I_\pm (\tau,\omega) =  \frac{1}{2}\left[G_{+}(\tau,\omega)
\pm G_{-}(\tau,\omega) \right],
\label{Ipm}
\eeq
with
\begin{equation}
G_{\pm}(\tau,\omega) = \left[ \frac{1}{\beta_{\pm}}
\sinh\left(\beta_{\pm}\,\frac{\Lambda\tau}{2}\right)
+ \cosh \left(\beta_{\pm}\,\frac{\Lambda\tau}{2}\right) \right]
\,e^{- \frac{\Lambda\tau}{2}},
\label{Gpm}
\end{equation}
and
\bea
&&\beta_{\pm} = a_+  \pm \imath a_{-},\hspace{0.30cm}
a_{\pm} = \pm \frac{1}{\sqrt{2}} \left[\rho \pm \left(1- \frac{2p^2}{\Lambda}\right) \right]^{1/2},
\nn\\ &&
\rho^2= \left(1- \frac{2p^2}{\Lambda}\right)^2 + \left(\frac{2 B \omega}{\Lambda}\right)^2.
\label{a-rho}
\eea
Close inspection of Eq.~(\ref{Gpm}) reveals that convergence in the
$\tau \rightarrow \infty$ demands $a_+ < 1$, which implies the following
constraint on the value of the memory parameter $\Lambda$ :
\begin{equation}
\Lambda > \frac{B^2\,\omega^2}{2\,(m\omega^2 + k)} \rightarrow \frac{B^2}{2m} .
\label{constraint}
\end{equation}
This convergence criterium may be compared to the one found in
Ref.~\refcite{menezes3}. Note that, because $\beta_\pm$ is complex, the
non-Markovian evolution is also oscillatory. However, as will be discussed
in the next section, for $\Lambda \neq \infty$, the effective non-Markovian
oscillation frequency can be significantly smaller that the corresponding
Markovian frequency.

Next, we consider the two-point correlation function - we follow the
derivation strategy developed in Ref.~\refcite{menezes2}. From Eqs.~(\ref{nmeta})
and (\ref{nmq}), we have
\begin{eqnarray}
\langle q_i(\tau,\omega)q_j(\tau',\omega')\rangle_{\eta}
= 4\pi \delta(\omega + \omega') \, \Delta_{ij}(\tau,\omega;\tau'\omega'),
\label{nmcorr}
\end{eqnarray}
where
\beq
\Delta_{ij}(\tau,\omega;\tau',\omega') = \int_{0}^{\tau}\,d\tau_1\,\int_{0}^{\tau'}\,d\tau_2\,
\Gamma_{im}(\tau-\tau_1,\omega) \, \Gamma_{mj}(\tau'- \tau_2,\omega) \,
M_\Lambda(|\tau_1-\tau_2|) .
\eeq
Using double Laplace transformations of $\Delta_{ij}(\tau,\omega;\tau'\omega')$,
one obtains
\bea
\Delta_{ij}(z,\omega;z',\omega') &=& \int_0^\infty d\tau \ e^{-z\tau} \ \int_0^\infty d\tau' \
e^{-z'\tau'} \Delta_{ij}(\tau,\omega;\tau',\omega')  \nn \\
&=& \Gamma_{im}(z, \omega) \, \Gamma_{mj}(z',\omega) \,
\left[\frac{M(z)+M(z')}{z+z'}\right].
\label{15nova}
\eea
Now, using Eq.~(\ref{gamma}) to eliminate $M(z)$ and $M(z')$, one can write
\beq
\Delta_{ij}(z,\omega;z',\omega') = \left[\frac{\Gamma_{il}(z,\omega)
+ \Gamma_{il}(z',\omega)}{z+z'}
- \Gamma_{im}(z,\omega)\Gamma_{ml}(z',\omega)\right]
D_{lj}(\omega),
\label{16nova}
\eeq
Using double inverse Laplace transformations in this equation, leads to the
non-Markovian two-point correlation function
\begin{eqnarray}
\langle q_i(\tau,\omega)q_j(\tau',\omega')\rangle_{\eta} =
4\pi\delta(\omega + \omega')
\left[\Gamma(|\tau' - \tau|,\omega)
- \Gamma(\tau,\omega)\Gamma(\tau',\omega)\right]_{il} D_{lj}(\omega).
\label{17nova}
\end{eqnarray}
Employing Eq.~(\ref{nmg2p}) for $\Gamma$ in this expression, gives the complete
and explicit solution for the non-Markovian two-point correlation function
in Fourier space.

It is not difficult to show that the large-$\tau$ limit of the non-Markovian two-point
correlation function is given by
\begin{eqnarray}
\lim_{\tau \rightarrow \infty} \, \langle q_i(\tau,\omega)q_j(\tau,\omega')\rangle_{\eta}
= \frac{2\pi\delta(\omega + \omega')}{p^4 + \omega^2B^2}
\, (\delta_{ij}p^2- \omega B \varepsilon_{ij}).
\label{nmcorrf}
\end{eqnarray}
As expected physically, the asymptotic limit is the same as in the Markovian case.
Differences arise at finite $\tau$. In particular, the memory kernel implies a slower
convergence toward equilibrium, but with a less oscillatory Green's function then
the corresponding Markovian one, as will be shown by an explicit numerical calculation
in the next section.

%
\section{Evolution towards equilibrium - numerical results}
\label{sec:equil}

In the present section we explore the qualitative differences between the evolution toward
equilibrium of the Markovian and non-Markovian processes for the present complex-action
problem. As argued previously, the introduction of a memory kernel can be helpful with
the requirements of small-step Langevin times in a numerical simulation of
complex-action Langevin equations. Specifically, we will show that the Green's function
of the non-Markovian process is less oscillatory in $\tau$ than the one corresponding
to the Markovian process with the same model parameters.

We consider first Markovian evolution. The matrix-valued retarded Green's function has
elements $g_{ij}(\tau,\omega)$ given in Eq.~(\ref{mg}). For our purposes, it is sufficient
to consider just one of its entries, since $g_{11}(\tau, \omega) = g_{22}(\tau, \omega)$,
$g_{12}(\tau, \omega)$  has the same qualitative behavior as $g_{11}(\tau, \omega)$, and
$g_{12}(\tau, \omega) = -g_{21}(\tau,\omega)$. Therefore, let us focus on
$g_{11}(\tau,\omega)$. In Fig.~\ref{1} we plot $g_{11}(\tau,\omega)$ for a specific
value of $\omega$ and arbitrarily chosen values of $k,m$ and $B$ - the value of $B$ is
intentionally chosen somewhat larger than other parameters in Planck units to highlight
more clearly the oscillatory character of the Green's function. The convergence of the
Markovian process is clearly seen in the Fig.~\ref{1}, as well as its oscillatory behavior.
From Eq.~(\ref{mg}), it should be clear that for non-zero values of $k$ the convergence to
equilibrium is faster than for $k=0$ case.
\begin{figure}[htb]
\begin{center}
\includegraphics[height=90mm,width=90mm]{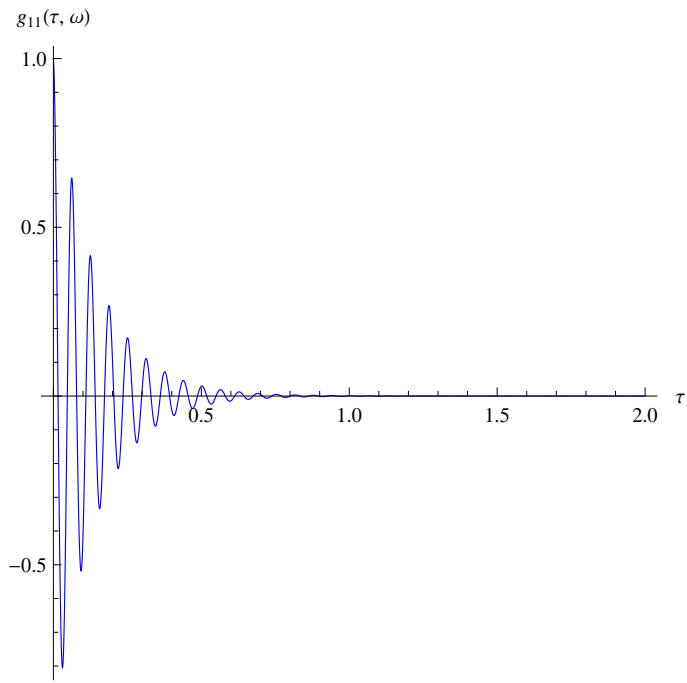}
\caption{Markovian retarded Green's function $g_{11}(\tau,\omega)$ for $\omega = 2$. Parameter
values are $k = 3$, $m = 1$, and $B = 50$ in Planck units.}
\label{1}
\end{center}
\end{figure}

Next, using the same values of $k,m$ and $B$, we examine the non-Markovian retarded
Green's function, $G(\tau, \omega) =\Gamma(\tau,\omega)\theta(\tau)$,
with the matrix $\Gamma$ given by Eq.~(\ref{nmg2p}). Similarly to the Markovian case,
we have $G_{11}(\tau,\omega) = G_{22}(\tau, \omega)$, $G_{12}(\tau, \omega) = -
G_{21}(\tau, \omega)$ and $G_{21}$ has the same qualitative behavior as $G_{11}$.
Therefore let us consider $G_{11}(\tau, \omega)$; its $\tau$ dependence is
illustrated in Figure~\ref{2}. The figure clearly shows that the pattern of oscillations is much
broader than the Markovian counterpart in Fig.~\ref{1}. That is, the oscillations are of
lower frequency. We mention also that similarly to the Markovian case, for non-zero
values of $k$ we notice that the convergence to equilibrium is faster than the $k=0$
case, even though this conclusion is not obvious for NMSQ.

\begin{figure}[htb]
\begin{center}
\includegraphics[height=90mm,width=90mm]{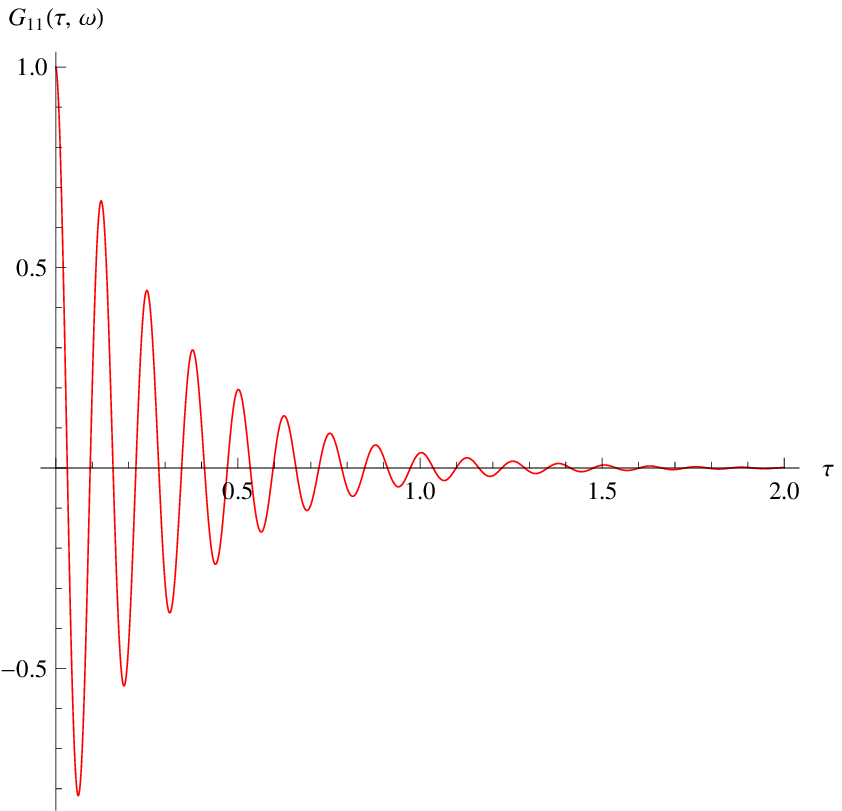}
\caption{Non-Markovian retarded Green's function $G_{11}$ for the diffusion
problem. The values used were the same as the previous figure with a memory parameter
$\Lambda = 10000$ in Planck units.}
\label{2}
\end{center}
\end{figure}

An important feature of the non-Markovian process, at least in the context of the present
model, is that the lowering of the oscillation frequency saturates for some value of the
memory parameter~$\Lambda$ - smaller values of $\Lambda$ do not decrease the frequency of
oscillations, they only retard more the evolution toward equilibrium. Therefore, there is
a compromise between lowering of the oscillation frequency and time of equilibration that
has to be verified case by case to benefit from a non-Markovian stochastic quantization
in a real, large scale numerical simulation.

Although our results are for a specific example, it should be clear that the introduction
of a memory kernel will have a similar effect in other situations. This is so because on general
physical grounds, a memory kernel as introduced here has the effect of delaying equilibration
and hence slowing down eventual oscillations in the time evolution - examples in other
contexts can be found in Refs.~\refcite{{mem1},{mem2}}. The obvious consequence
for large-scale numerical simulations is that the Langevin time evolution can be sampled with
with larger step-sizes. Evidently, delayed equilibration has computational costs. However,
such costs might be a price to be payed for smoother time evolution.

%
\section{Conclusions and Perspectives}
\label{sec:concl}

In the present paper we studied qualitative differences between Markovian and non-Markovian
stochastic quantization in a simple model with a complex action - a topological quantum
mechanical action which is analog to a Maxwell-Chern-Simons action in the Weyl gauge.
Complex actions introduce oscillations in the Langevin time evolution toward equilibrium.
Such oscillations can introduce difficulties in numerical simulations, as the requirement
of short Langevin time-steps for achieving convergence to the correct equilibrium state. The
introduction of a memory kernel in the Langevin equation has the effect of delaying
equilibration which in turn slow down oscillations in the time evolution. The practical
consequence for large-scale numerical simulations is that the Langevin time evolution
can be sampled with with larger step-sizes.

As we remarked in the previous section, although our results are for a specific example,
the effect of softening oscillations is a generic physical feature of memory kernels
and because of this one expects that non-Markovian stochastic quantization might
be helpful for other, more complicated problems with complex actions.

\section{Acknowledgments}
Work partially financed by CNPq and FAPESP (Brazilian agencies).

%
\appendix
\label{sec:app}

\section{Calculation of the non-Markovian Green's function}

In this Appendix we outline the derivation of the inverse
Laplace transform for the $\Gamma$ matrix. Our derivation is for a memory
kernel $M_{\Lambda}(\tau)$ of exponential form:
\beq
M_\Lambda(\tau) = \frac{1}{2} \, \Lambda \, e^{-\Lambda \mid\tau\mid}\,,
\label{mem}
\eeq
whose Laplace transform is
\beq
M_\Lambda(z)=\int_{0}^{\infty}d\tau\,M_{\Lambda}(\tau)\,e^{-z\tau}
= \frac{1}{2}\,\frac{\Lambda}{z + \Lambda} .
\label{Mz}
\eeq
From Eq.~(\ref{gamma}), one has that $\Gamma_{ij}(z,\omega)$ can be written as
\begin{equation}
\Gamma_{ij}(z,\omega) = \frac{\delta_{ij} \, \left[ z + p^2 M_\Lambda(z)\right] -
\varepsilon_{ij}\,\omega\,B M_\Lambda(z)}{\left[z+p^2 M_\Lambda(z)\right]^2
+ \omega^2 B M^2_\Lambda(z)} .
\label{33}
\end{equation}
In order to obtain the inverse Laplace transform of this equation, one needs to find
the zeros of the denominator. Expanding the denominator using the explicit form of the
memory kernel of  Eq.~(\ref{Mz}), one has to find the zeros of the quartic equation
\beq
z^4 + 2\Lambda z^3 + (\Lambda^2 +
p^2 \Lambda) \, z^2 + p^2 \Lambda^2 \,z + (p^2 + B^2 \omega^2) \Lambda^2/4 = 0.
\label{41}
\end{equation}
The four roots are given by
\begin{eqnarray}
z_1 &=& - \frac{1}{2}\left[\Lambda + \left(a_+ + \imath a_-\right)\right],\hspace{0.5cm}
z_2 = - \frac{1}{2}\left[\Lambda - \left(a_+ + \imath a_-\right)\right], \nn \\
z_3 &=& -\frac{1}{2}\left[\Lambda + \left(a_+ - \imath a_-\right)\right], \hspace{0.5cm}
z_4 = -\frac{1}{2}\left[\Lambda - \left(a_+ - \imath a_-\right)\right]
\label{roots}
\end{eqnarray}
where the $a_\pm$ are given in Eq.~(\ref{a-rho}). Obtaining the inverse Laplace transformation with a denominator as
$(z-z_1)(z-z_2)(z-z_3)(z-z_4)$ is a straightforward, albeit tedious procedure. The final result for $\Gamma_{ij}(\tau,\omega)$
can be written in the form presented in Eqs.~(\ref{nmg2p})-(\ref{a-rho}).

\end{document}